\documentclass[aps,pra,superscriptaddress,twocolumn,notitlepage,bibnotes]{revtex4-1}
\usepackage{graphicx}
\usepackage{color,pgfplots}
\usepackage{epstopdf}
\usepackage{braket}
\usepackage{amsmath}
\usepackage{caption}
\usepackage{notes2bib}
\usepackage{float}
\usepackage{latexsym}
\usepackage{braket}
\usepackage{pdfpages}

\usepackage[normalem]{ulem}

\usetikzlibrary{plotmarks}
\pgfplotsset{compat=1.10}

\newcommand{\im}{\ensuremath{ \text{i} }}

\makeatletter
\AtBeginDocument{\let\LS@rot\@undefined}
\makeatother

\begin{document}

\title{Perfect Chirality with imperfect polarisation}

\author{Ben Lang}
\email{Current address: School of Physics and Astronomy, University
of Nottingham, UK}
\affiliation{Quantum Engineering Technology Labs, H. H. Wills Physics Laboratory and Department of Electrical \& Electronic Engineering, University of Bristol, BS8 1FD, UK}

\author{Dara  P. S. McCutcheon}
\affiliation{Quantum Engineering Technology Labs, H. H. Wills Physics Laboratory and Department of Electrical \& Electronic Engineering, University of Bristol, BS8 1FD, UK}

\author{Edmund Harbord}
\affiliation{Quantum Engineering Technology Labs, H. H. Wills Physics Laboratory and Department of Electrical \& Electronic Engineering, University of Bristol, BS8 1FD, UK}

\author{Andrew B. Young}
\affiliation{Quantum Engineering Technology Labs, H. H. Wills Physics Laboratory and Department of Electrical \& Electronic Engineering, University of Bristol, BS8 1FD, UK}

\author{Ruth Oulton}
\affiliation{Quantum Engineering Technology Labs, H. H. Wills Physics Laboratory and Department of Electrical \& Electronic Engineering, University of Bristol, BS8 1FD, UK}

\begin{abstract}
Unidirectional (chiral) emission of light from a circular dipole emitter into a waveguide is only possible at points of perfect circular polarisation (C points), with elliptical polarisations yielding a lower directional contrast. However, there is no need to restrict engineered systems to circular dipoles and with an appropriate choice of dipole unidirectional emission is possible for any elliptical polarization. Using elliptical dipoles, rather than circular, typically increases the size of the area suitable for chiral interactions (in an exemplary mode by a factor $\sim 30$), while simultaneously increasing coupling efficiencies. We propose illustrative schemes to engineer the necessary elliptical transitions in both atomic systems and quantum dots.
\end{abstract}


{\let\newpage\relax\maketitle}

\emph{Introduction} - Nanostructures are often employed to finely control light. A common application is confining light into narrow channels to maximise the interaction probability between photons and a matter system, such an atom or quantum dot (QD) located at the focus point, a situation sometimes called the ``1D atom'' \cite{1D_atom, atom_in_1d_cavity, vuckovic_drop_filter}. The light propagating through these narrow channels has components of transversely rotating (``rolling'') electric fields \cite{p_wheel}, a consequence of Gauss' law \cite{Rici_coles1}. This rolling polarisation can give rise to \emph{chirality}, a near-field effect where atomic transitions described by circular dipoles radiate in a preferred direction \cite{Young_PRL, spin_hall_light, Unidirectional_surface_waves, Ryan_Collective}.

These chiral behaviours have been recognised as a new tool in the development of light-matter interfaces \cite{chiral_quantum_optics}. One application is constructing on-chip quantum memories with charged QDs, the qubit-states of which possess oppositely handed circular dipoles. Without chirality distinguishing these dipoles in-plane is cumbersome: requiring the collection and interference of beams propagating in orthogonal directions \cite{cant_do_in_plane}.

Perfect chirality (100\% emission in a single direction) is frequently pursued by attempting to place the emitter at a point of perfect circular polarisation (a C point). These points are scarce in most structures. In nanofibre based waveguides only elliptical polarisation is practically accessible \cite{Rauschenbeutel}, while nano-beam and photonic crystal structures support circular polarisation at a few accessible locations, but the light field is elliptically polarised over the majority of the mode volume \cite{Kobus_c_points, Rici_coles1, Lodahl1}. 

However, in the typical case of elliptical polarisation perfect chiral behaviour is still possible given the correct transition dipole \cite{Rauschenbeutel, Unidirectional_surface_waves,Francisco_2021, Francisco_2018, VanMechelen_16, Francisco_2014}. This suggests the alternative strategy of engineering the emitter dipole, the topic of this letter. This approach is attractive for quantum light-matter interfaces as higher coupling efficiencies will typically be possible using elliptical polarisation.

\emph{Emission} - We begin with a 1D waveguide supporting a single forward and single backward propagating mode described by classical, complex electric fields given by $\mathbf{E}_f(\mathbf{r})$ and $\mathbf{E}_b(\mathbf{r})$. Light is emitted by a matter system (MS), which is assumed to be a two level quantum system with energy levels connected by an optical dipole transition with dipole moment $\mathbf{d}$. The MS could represent an atom or QD for example. It is placed at a location in the waveguide, $\mathbf{r}$, and interacts with the electric fields at this location.

We assume that initially the MS is in its excited state with no photons in the waveguide modes. From Fermi's golden rule \cite{supplementary}, the likelihood of the MS decaying via the emission of a photon in the forwards direction is proportional to $|\mathbf{d}^*\cdot \mathbf{E}_f|^2$ while for the backwards direction it is $|\mathbf{d}^*\cdot \mathbf{E}_b|^2$. As the forwards and backwards modes are related by time-reversal symmetry, $\mathbf{E}_f = \mathbf{E}_{b}^*$, directionality is usually associated with circular polarisation where the two propagation directions are orthogonally polarised.

\begin{figure}[t]
\begin{centering}
\includegraphics[scale=0.21]{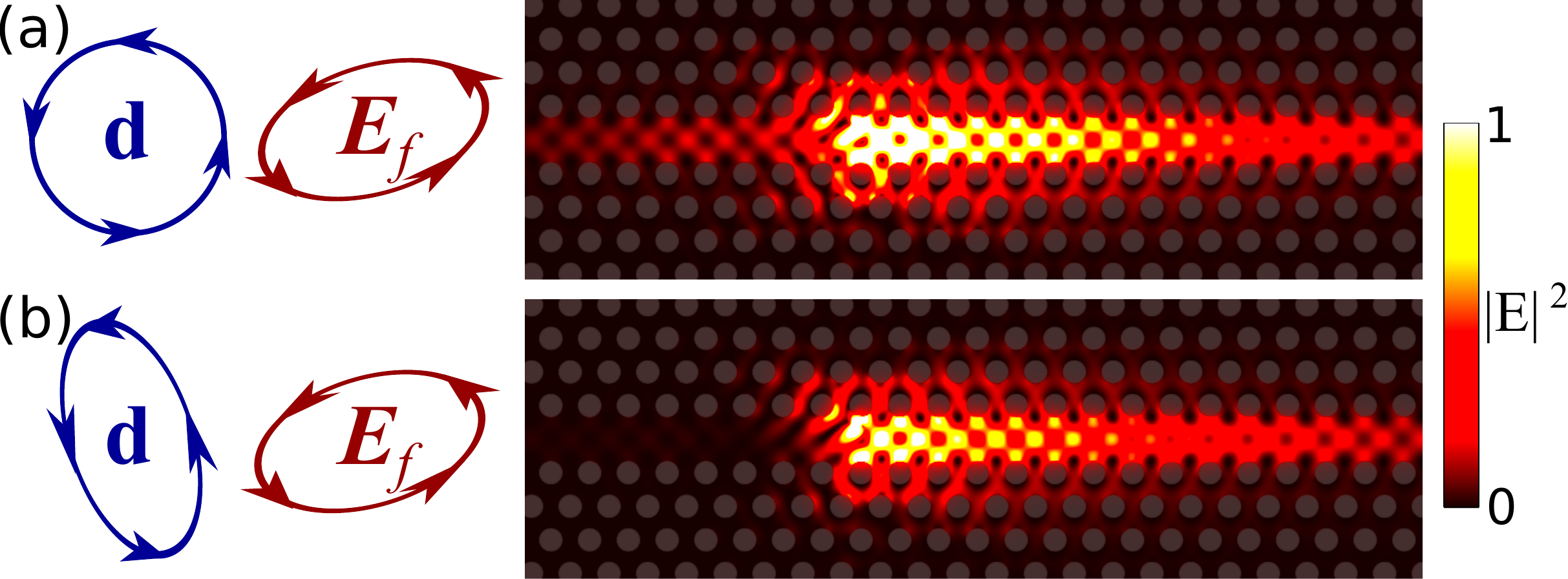}
\end{centering}
\caption{(a) Simulation of a circular dipole source located at a point of elliptical polarisation in a photonic crystal waveguide. Light is emitted in both directions but the forwards direction is preferred. (b) Replacing the circular dipole source with the depicted elliptical one results in unidirectional emission as this dipole is orthogonal to the polarisation of the backward mode.}
\label{FDTD}
\end{figure}

\begin{figure}
\includegraphics[scale=0.4]{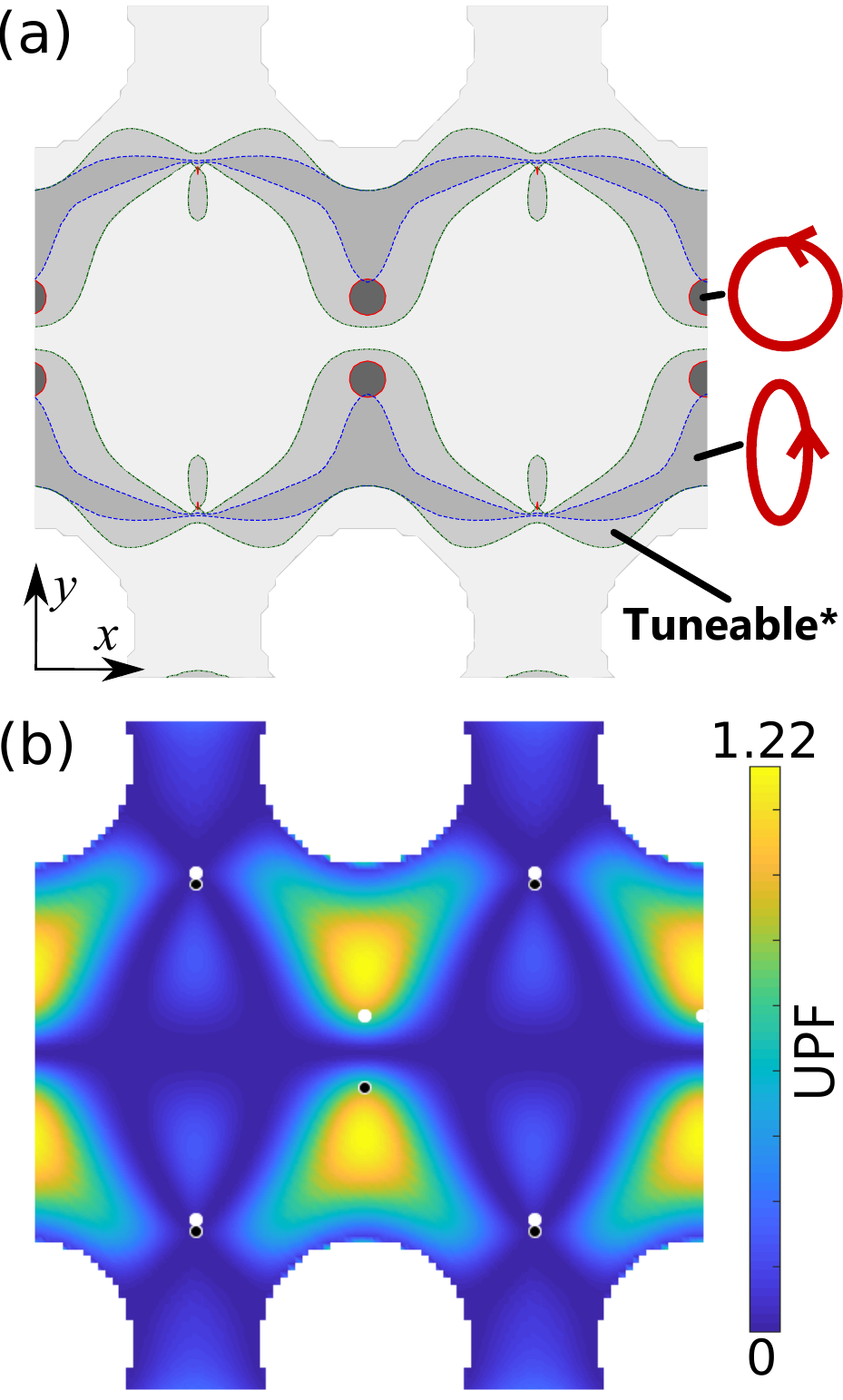}
\caption{(a) Positions where different types of dipole source have $|D|\geq 0.9$ in a waveguide mode. Darkest - circular dipole; next darkest - (fixed) elliptical dipole as shown ($S_1 = \sqrt{1/2}$). Second palest - Variable elliptical dipole, tuned for maximum $|D|$ subject to $S_3 \geq 1/\sqrt{2}$ (marked tuneable$^*$). Large white circles - air holes. (b) Unidirectional Purcell factor. White/black dots are right/left polarised C points where unidrectional emission occurs for circular dipoles. In contrast, all locations with UPF$\neq0$ enable unidirectional coupling with suitably elliptical dipoles.}
\label{map}
\end{figure}

When the local polarisation is elliptical, as depicted in Fig.\ref{FDTD}(a), these dot-product rules indicate that a circular dipole radiates in both directions in the waveguide, but with differing intensities. This is confirmed using a Finite Difference Time Domain simulation in which a circular dipole is placed in a photonic crystal waveguide at a point of elliptical polarisation, Fig.\ref{FDTD}(a)\cite{meep}\footnote{Simulation parameters. Structure as fig.\ref{map}. Source frequency $f = 0.2791 (c/a)$. Source position $(x,y) = (-0.125, 0.3958)a$ relative to origin centred on waveguide between two holes. In (a): $\mathbf{d} = 1/\sqrt{2}(1, \im)$, (b): $\mathbf{d} = (-0.250-0.448\im, 0.856-0.061\im)$.}.

However, there will always be a dipole orthogonal to the polarisation of the backwards mode, $\mathbf{d}_{\perp}^{*} \cdot \mathbf{E}_b = 0$, so that there is no emission backward. For any polarisation except linear, this dipole will be non-orthogonal to the polarisation of the forwards mode, $\mathbf{d}_{\perp}^{*} \cdot \mathbf{E}_f \neq 0$ resulting in emission in only the forward direction \cite{Unidirectional_surface_waves, Rauschenbeutel, Francisco_2021, Francisco_2018, VanMechelen_16, Francisco_2014}. An example is shown in Fig.\ref{FDTD}(b). By setting the dipole's long axis orthogonal to the long axis of the polarisation the linear components have been cancelled, leaving only a circular-like effect (unidirectional emission). This demonstrates that the directional contrast is not in general limited by the degree of circular polarisation, and can be unity with any polarisation (except exactly linear) given the correct emission dipole.

The chirality can be measured using the directional contrast, the difference between the power radiated forwards and backwards divided by the sum of the two, $D = ( |\mathbf{d}^{*} \cdot \mathbf{E}_f|^2 - |\mathbf{d}^{*} \cdot \mathbf{E}_b|^2 ) / (|\mathbf{d}^{*} \cdot \mathbf{E}_f|^2 + |\mathbf{d}^{*} \cdot \mathbf{E}_b|^2)$. For a circular dipole it takes the same value as the (normalised) Stokes parameter describing the degree of circular polarisation $D = S_3 =  2 \, \text{Im}(E_x^* E_y)/|\mathbf{E}|^2$ \cite{us_disorder}. In figure \ref{map}(a) we assess a typical photonic crystal waveguide mode with wavevector $k_x = 0.395 (2\pi / a)$, for lattice constant $a$ \cite{mpb}. The hole radii/slab height are $r=0.3a$ and $h=0.6a$ respectively as in \cite{optimal_us}. The polarisation varies spatially, such that a circular dipole is strongly directional ($|D|\geq0.9$) over small areas as indicated by the darkest shading. However, for each location (except on lines of zero area) there is a dipole that is unidirectional. A specific elliptical dipole is shown, which is ``half circular'' in the sense that $S_3 = S_1 = 1/\sqrt{2}$, with $S_1 = (|E_y|^2-|E_x|^2)/|\mathbf{E}|^2$ the Stokes parameter for rectilinear polarisation. It is noticeable that with this dipole a far larger area in the waveguide is useful for unidirectional coupling. Finally we mark the areas for which there exists a dipole that is at least half-circular ($S_3 \geq 1/\sqrt{2}$) which has $|D|\geq0.9$. This area is $\sim 30$ times larger than that in which $|D|>0.9$ occurs with a circular dipole. Had we considered $|D|>0.95$ the increase would instead be $\sim \times 45$ \cite{supplementary}.

The two crucial parameters for a chiral light-matter interface are the directional contrast and the fraction of light that is emitted into the waveguide, known as the Beta factor \cite{Scarpelli_2019}. We have shown that one can recover high directional contrast with elliptical polarisation. However, it is important to assess how this will effect the Beta factor, which is largely determined by the coupling rate between the MS and waveguide ($\propto |\mathbf{d}^*\cdot \mathbf{E}_f|^2 + |\mathbf{d}^*\cdot \mathbf{E}_b|^2$). Typically a higher electric field intensity will be possible away from a C point \cite{Scarpelli_2019, edge_us}. However, away from the C point the polarisations of the forwards and backwards modes are non-orthogonal, and thus the dipole orthogonal to the backward mode has poorer overlap with the forward mode. Accounting for both effects the overall unidirectional coupling strength varies as $S_3^2(\mathbf{r}) |\mathbf{E}(\mathbf{r})|^2 |\mathbf{d}|^2$.

In Fig.\ref{map}(b) we explore the impact of these competing effects. At each location the unidirectional coupling rate after the dipole has been adjusted for the new location is plotted. The rate is normalised to the emission rate in bulk GaAs (refractive index $n=3.45$) to give a Purcell factor; $P(\mathbf{r}) = S_3^2(\mathbf{r}) |\mathbf{E}(\mathbf{r})|^2  \,\, (3 / 8 \pi v_g f^2 n) $ with $f$ the mode frequency and $v_g$ the group velocity, expressed in units of  $c/a$ and $c$ respectively \cite{Stephen_Hughes1}. In Fig.\ref{map} $f= 0.262 (c/a)$ and $v_g = 0.03 c$.

Coupling matching that at a C point is achieved over a significant area, indeed coupling is not maximised at the C points: It is over 50\% higher at other locations where increased field intensity has more than compensated for the reduction in overlap between the dipole and the forwards mode. Comparison with part (a) of the figure further shows that the elliptical dipole not only couples unidirectionally in more places, but also does so in places with stronger coupling.

\emph{Scattering} - As well as emission, unidirectional coupling also has important consequences for the scattering of photons from the MS. To model scattering we assume that initially the MS is in its ground state and the forwards waveguide mode is populated with a single photon. The scattering amplitudes with which the MS (re)-directs the incident photon are then calculated using a method based on the photonic Green's function. Our aim is to avoid input/output theory which can cause confusion in chiral systems \footnote{For example \cite{One_way_mirror} and \cite{Lodahl1} draw differing conclusions using input/output theory.} The calculation involves a number of steps, summarised here and detailed in the supplementary information \cite{supplementary}.

The interaction between the MS and light is included perturbatively. The probability amplitude in the state $\ket{k}$, $\gamma_{k}(t)$ at time $t$ is given by $\gamma_{k}(t) = \sum_{l=0}^{\infty} \gamma_{k}^{(l)}(t)$, where $\gamma_k^{(0)}$ is defined by the initial condition and all subsequent orders are calculated from the previous according to: 

\begin{equation}
\gamma_{k}^{(z+1)}(T) =  \int \limits^{T} \sum_m \bra{k} \hat{H} \ket{m} \frac{ \gamma_{m}^{(z)}(t)}{\im \hbar} e^{\im(E_k - E_m) t /\hbar} dt,
\end{equation}
\noindent
with the sum over all states of the system, $E_m$ the unperturbed energy of the state $\ket{m}$, and $H$ the interaction Hamiltonian \cite{Grynberg_textbook}.

Adopting the long photon limit (a single frequency photon), we can expand the integration range to $\pm \infty$. Further the Hamiltonian is assumed to turn off slowly as one approaches $|t| \rightarrow \infty$ as in \cite{Shen_Fan_Perturb}. Together these assumptions result in each even perturbation order being given by the previous even order times a fixed multiplier, allowing all orders to be collected using a geometric series.

The Hamiltonian for light coupling to a two level system with a single optical transition (in the rotating wave approximation) is given by: $\hat{H} = - \im \hat{\sigma}^{+} \hat{l}  + \text{h.c.}$ with \cite{welsch2, stephen_hughes_mollow_triplet_metalic_np},

\begin{equation}
\hat{l} = \mathbf{d} \iint \textbf{G}(\mathbf{r}, \mathbf{r}', \omega') \sqrt{ \frac{\hbar \text{Im}[\epsilon(\mathbf{r}',\omega')]}{\epsilon_{0} \pi}} \mathbf{\hat{f}}(\mathbf{r}',\omega') d^{3}\mathbf{r}' d\omega'\,,
\label{Rotating_Hamiltonian}
\end{equation}
where $\textbf{G}(\mathbf{r}, \mathbf{r}', \omega')$ is the tensor electromagnetic Green's function connecting locations $\mathbf{r}$ and $\mathbf{r}'$ at frequency $\omega'$. The raising operator of the two-level system is $\hat{\sigma}^{+}$. The dielectric profile is given by $\epsilon$ and $\mathbf{\hat{f}}(\mathbf{r}',\omega')$ is the annihilation operator of a bosonic excitation in the dielectric material and its associated electromagnetic field \cite{green_bible}. 

$\hat{H}$ is inserted into the perturbation model and standard Green's function identities \cite{welsch1, stephen_hughes_mollow_triplet_metalic_np, SH_P_value}, are exploited to simplify the spatial and frequency integrals. Finally an integration over a small frequency window is introduced, representing the resolution of a detector. This is necessary to correctly normalise the density of states.

The result of this calculation is the following equations for the (complex) transmission and reflection amplitudes (the probability amplitudes with which the photon will be found in the forwards/backwards modes in the long-time limit):

\begin{equation}
t = 1 - \frac{\mathbf{d} \cdot  \mathbf{E}_{f}^*(\mathbf{r}) \,\, \mathbf{d}^* \cdot  \mathbf{E}_f(\mathbf{r})}{D} \,,
\label{transmission}
\end{equation}
\begin{equation}
r = -\frac{\mathbf{d} \cdot  \mathbf{E}_{b}^*(\mathbf{r}) \,\, \mathbf{d}^* \cdot  \mathbf{E}_f(\mathbf{r})}{D} \,,
\label{reflection}
\end{equation}

\begin{equation}
\begin{split}
D = &\frac12 \left[ |\mathbf{d}^* \cdot  \mathbf{E}_{f}(\mathbf{r})|^2  + |\mathbf{d}^* \cdot \mathbf{E}_{b}(\mathbf{r})|^2  \right] \\
&+ \zeta \left(\mathbf{d} \cdot  \mathbf{G}_{\text{Loss}} \cdot  \mathbf{d}^* + \im \hbar\epsilon_0\Delta \right)\,,
\end{split}
\end{equation}
where $\mathbf{G}_{\text{Loss}}$ represents the Green's function of the loss mode(s) (with both spatial dependencies set to $\mathbf{r}$) and $\Delta$ the detuning between the photon and the transition frequency. The $\mathbf{E}$ terms are the electric fields of Bloch modes normalised as $\int \epsilon(\mathbf{r}) |\mathbf{E}(\mathbf{r}) |^2 = 1$ with the integration over a single unit cell of the waveguide (or over the cross section for a translationally invariant waveguide like a fibre). In a translationally periodic (invariant) waveguide $\zeta = \frac{2 v_g}{a \omega}$ ($\zeta = \frac{2 v_g}{\omega}$) with $\omega$ the transition angular frequency.  

The derivation can be extended to systems with more than two levels and multiple dipole-allowed transitions. Consider a four level system with only two allowed transitions, one connecting $\ket{g_1}$ to $\ket{e_1}$, and the other $\ket{g_2}$ to $\ket{e_2}$, an arrangement we denote as $\emph{II}$ by analogy to the well known $\Lambda$ and $V$ systems \cite{atom_switch}(Fig.\ref{II_system}). Here there are two dipoles, $\mathbf{d}_1$, $\mathbf{d}_2$, one for each transition. Similarly there are two detunings. If the system is initially in one of the ground states then equations (\ref{transmission}, \ref{reflection}) apply, using the detuning and dipole associated with the transition available to the initial ground state. An initial superposition of ground states simply implies a superposition of reflection/transmission coefficients:

\begin{equation}
\begin{split}
\large(\alpha \ket{g_1} + \beta \ket{g_2} \large) \ket{1_f}&  \rightarrow \\
 \alpha \ket{g_1} ( t_1\ket{1_f} + r_1\ket{1_b} )&  + \beta \ket{g_2}( t_2 \ket{1_f} + r_2 \ket{1_b} ),
\end{split}
\end{equation}
where $r_n$, $t_n$ are the reflection/transmission coefficients calculated from equations (\ref{transmission}, \ref{reflection}) using the dipole and detuning of the $n^{\text{th}}$ transition. This is the underlying mechanism behind some proposals to entangle the emitter with a photon \cite{Young_PRL}.

Scattering calculations for systems where a single ground/excited state has multiple allowed transitions, such as \emph{V} and $\Lambda$ arrangements require a more complicated treatment \cite{Hughes_Anisotropy_interference, Bens_thesis}. However, the dot-product rules that determine directionality are unchanged.

Chiral interactions at C points are characterised not just by an excited MS radiating light in only one direction, but also by a single-photon transmission coefficient that approaches $-1$ for low loss - i.e. transmission of the incident photon with a phase shift of $\pi$. This phase shift is exploited in several proposals for quantum information circuits \cite{Young_PRL, chiral_component, Lodahl1, exploiting_lambda_sys}. However this effect is not fundamentally associated with either circular dipoles or polarisations, but emerges whenever the interaction between the MS and waveguide is unidirectional. 

\begin{figure}
\includegraphics[scale=0.62]{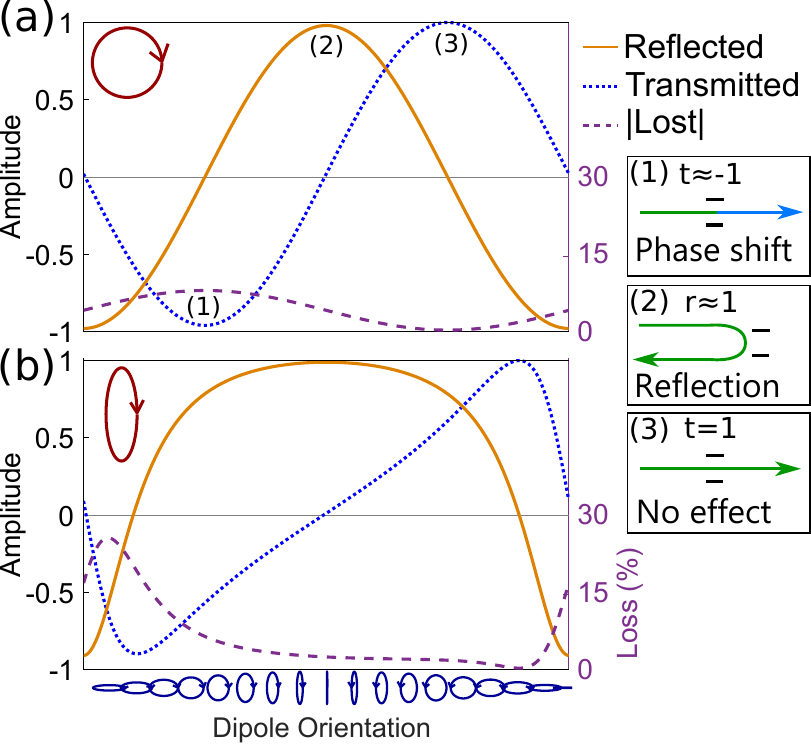}
\caption{Single photon reflection (solid) and transmission (dotted) coefficients for a two level MS as a function of the dipole moment. Curves are calculated using equations (\ref{transmission}, \ref{reflection}) with $\Delta$ = 0, $|\mathbf{E}| = |\mathbf{d}| = 1$ and $\zeta \mathbf{d} \cdot  \mathbf{G}_{\text{Loss}} \cdot  \mathbf{d}^* = 0.01$. Dashed: lost intensity, right axis. In (a) the polarisation is circular so that a dipole matching the helicity of the light phase shifts a passing photon (1), while a dipole of opposite helicity transmits the photon with no phase shift (3). Linear dipoles reflect the photon (2). In (b) with elliptical polarisation note that there are still dipoles that transmit a photon with or without a phase shift.}
\label{rt}
\end{figure}

This is shown in Fig.\ref{rt} where the reflection and transmission coefficients from equations (\ref{transmission}, \ref{reflection}) are plotted. In parts (a) and  (b) the local polarisation is given by $\mathbf{E}_f  = 1/\sqrt{2} \begin{pmatrix} 1 & \im \end{pmatrix} $ and $\mathbf{E}_f  = 1/\sqrt{10} \begin{pmatrix} 1 & 3\im \end{pmatrix}$ (an arbitrary choice) respectively while along the $x$ axis the dipole is varied as $\mathbf{d}= \begin{pmatrix} \cos\theta & \im \sin \theta \end{pmatrix} $ with $\theta$ running from $0$ to $\pi$. A phase-shift ($t\approx -1$) occurs when $\mathbf{d}^*\cdot \mathbf{E}_b = 0$, a consequence of the fact that in this configuration the forwards/backwards emission rates are identical to those with a circular dipole at a C point, indeed there is no special feature that separates chirality with circles from that with ellipses.

More generally, many interesting effects have been predicted in theoretical frameworks where particular forward/backward emission rates are assumed \cite{chiral_component, exploiting_lambda_sys, dimer_formation, Lodahl1, atom_switch, multi_dimer_formation, multiple_atoms_gausian_pulse}, rather than specifying the dipoles/polarisations that determine these rates. The predictions of these works apply to all polarisation/dipole combinations that produce directionality.

Comparing the points with transmission approaching $-1$ of parts (a) and (b) of Fig.\ref{rt} notice the losses are higher in (b). Here the poorer dot product between the dipole and the forwards mode, combined with our assumption that $|\mathbf{E}|$ is unchanged between parts (a, b) has led to a reduced Beta factor. However, as discussed previously $|\mathbf{E}|$ can be much higher at elliptical points and this will often more than compensate for the lower overlap. As seen in Fig.\ref{map} the choices that maximise coupling (and consequently minimise this loss) are ellipses.

Some proposals require a MS with more than two levels. We consider two specific schemes, both with potential applications in quantum information. First that of \cite{Young_PRL}, which makes use of charged QDs, described as a \emph{II}-like four-level system. Second \cite{atom_switch}, with Caesium atoms described with three levels in a $\Lambda$-like configuration. In both cases there are two relevant optical transitions and in both ideally one transition couples only to the forwards direction, while the other couples only backwards, depicted in Fig.\ref{II_system}. As seen in part (b) of the figure the ideal pair consists of two elliptical dipoles, identical in all respects except for helicity (the arrowhead direction) which is opposite between them. That is, ideally $\mathbf{d}_1 = \mathbf{d}_2^*$ and $\mathbf{d}_1^* \cdot \mathbf{E}_b = 0$.

The non-orthogonality of the two dipoles in fig.\ref{II_system}(b) is no impediment to the quantum information proposals, as the orthogonality of dipoles in real space does not equate to orthogonality of quantum states in Hilbert space \cite{supplementary}.

\begin{figure}
\includegraphics[scale=0.22]{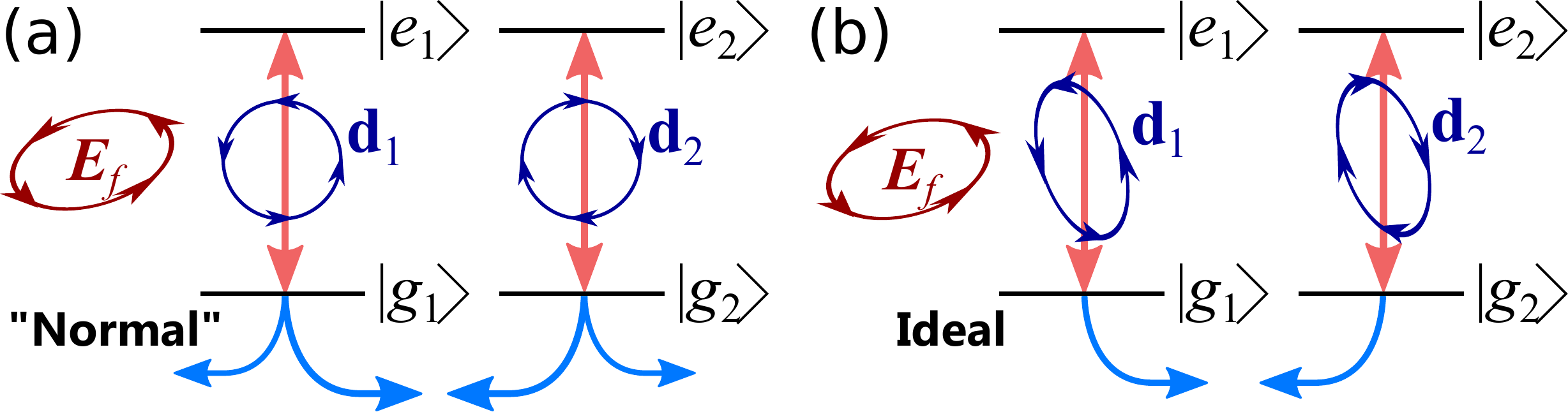}
\caption{(a) Four level system with two transitions with opposite circular dipole moments. Each transition interacts with both waveguide directions (blue arrows). (b) The dipole arrangement such that each interacts with only a single waveguide direction. For this polarisation (b) represents the ideal situation for the protocols discussed in the text, in contrast to (a).
}
\label{II_system}
\end{figure}

\emph{Dipole engineering} - Finally we propose illustrative schemes to achieve the necessary dipole engineering in either an atomic or a QD system, focusing on proposals that call for two transitions which are unidirectional in opposite directions \cite{atom_switch, Young_PRL}.

Atoms - Given a system with a circular transition dipole one can imagine rotating the system in 3D space so that the projection of the circular dipole onto the 2D plane spanned by the waveguide mode's electric field resembles the desired ellipse. If provided a system with oppositely circular transitions rotating it to make one transition couple only to the forward mode will result in the other coupling only to the backward mode. With atoms in vacuum \cite{atom-fibre, atom_switch}, it is the external magnetic field that defines the quantisation axis, so that the relevant transitions have circular dipole moments in the plane orthogonal to the magnetic field direction \cite{Caesium_atoms_quant_dipole, Grynberg_textbook}. Consequently tilting the applied magnetic field will have the desired effect.

QDs - In QDs there will be in-plane strain, which leads to mixing between the heavy and light holes. The dipole associated with the recombination of an electron with a light-hole has the opposite sense of circular polarisation to that for recombination with a heavy hole. As a result recombination with the mixed holes in the QD is related to an elliptical dipole \cite{Hole_mixing}, typically with 1\%-20\% degree of linear polarization ($\sqrt{1-S_3^2} = 0.01 \text{--} 0.2$). The dipoles of the two QD transitions are stretched along the same axis, providing the ideal configuration. The degree of linear polarisation in these dipoles can be enhanced to up to 40\% by annealing \cite{Harbord_2013}, and can be tuned $\pm$20\% with application of strain \cite{Valence_Band_tuning_strain}. Two strategies emerge: first, annealing allows the creation of QD-waveguide samples where the (randomly located) QDs are more likely to have a high directionality and stronger waveguide coupling; second, it may be possible to exploit strain-tuning techniques to modify the dipole of a particular QD \emph{in situ} to maximise the directionality.

\emph{Conclusion} - We propose the engineering of elliptical dipoles in quantum emitters as an approach to building chiral interfaces. These strategies offer the two-fold advantage of making far more of the space within a waveguide useful for directional interactions while simultaneously enabling a higher photon collection efficiency.

Advanced proposals call for a system with two transitions, each unidirectional but in opposite directions. This requires that the opposite circular dipoles are replaced with ellipses stretched along a shared axis. We have outlined methods to achieve these arrangements in both atomic and QD systems.

In summary, circular polarisations and transition dipoles are not necessary for chiral interactions, furthermore they are typically not the most efficient choices.

\vspace{0.5pc}

Additional references \cite{Loudon_textbook, denmark_mode_index, Hughes_laser_review} are used in the Supplemental information. 

Acknowledgements - We gratefully acknowledge EPSRC funding through the projects 1D QED (EP/N003381/1), SPIN SPACE (EP/M024156/1) and BL's studentship (DTA-1407622). This work was also supported by Leverhulme Trust Research Project Grant No. RPG-2018-213. We thank Joseph Lennon for help with data access issues.

\bibliographystyle{apsrev4-1}
\bibliography{bibliogrpahy}

\newpage
\clearpage


\section*{Supplementary material}

# I. DETAILED DERIVATIONS

## A. Scattering

This section presents the details of the calculation of the single-photon scattering amplitudes from a two-level matter system (MS).

We will consider the situation where, initially, a single photon is propagating in the forwards direction in the waveguide and the MS is in its ground state. To find the scattering equation we will assume the MS and light field are initially uncoupled, and the coupling between them will be included perturbatively.

We split our Hamiltonian into two parts; a non-interacting part that captures the behaviour of the isolated MS and light field ($\hat{H}_0$), and an interacting part that couples them ($\hat{H}_I$). The non-interacting part is assumed to support eigenmodes with known energies, but its exact form is not important. Making the rotating wave approximation the interacting part of the Hamiltonian is given by [1, 2]:

$$\begin{aligned}\hat{H}_I = &-\mathrm{i}\hat{\sigma}^+\mathbf{d}\cdot\iint \mathbf{G}(\mathbf{r},\mathbf{r}',\omega')\sqrt{\frac{\hbar\mathrm{Im}[\epsilon(\mathbf{r}',\omega')]}{\epsilon_0\pi}}\hat{\mathbf{f}}(\mathbf{r}',\omega')d^3\mathbf{r}'d\omega'\\ &-\mathrm{i}\hat{\sigma}^-\mathbf{d}^*\cdot\iint \mathbf{G}^*(\mathbf{r},\mathbf{r}',\omega')\sqrt{\frac{\hbar\mathrm{Im}[\epsilon(\mathbf{r}',\omega')]}{\epsilon_0\pi}}\hat{\mathbf{f}}^\dagger(\mathbf{r}',\omega')d^3\mathbf{r}'d\omega'.\end{aligned} \quad (1)$$

where $\mathbf{G}(\mathbf{r},\mathbf{r}',\omega')$ is the tensor Green's function connecting locations $\mathbf{r}$ and $\mathbf{r}'$ at frequency $\omega'$. $\mathbf{d}$ is the dipole moment of the transition, $\mathbf{r}$ the MS location and $\hat{\sigma}^\pm$ are the MS raising/lowering operators. Within the dielectric material there exist material degrees of freedom (for example electrons being displaced from their atomic nuclei) that interact with the electromagnetic field (this interaction is the source of the refractive index). $\hat{\mathbf{f}}/\hat{\mathbf{f}}^\dagger$ are bosonic field annihilation/creation operators that represent basic excitations of these material degrees of freedom and the related electromagnetic field [3]. $\epsilon$ is the relative electric permittivity.

Given an initial state of the system, described by a superposition of states selected from the spectrum of the non-interacting Hamiltonian with complex amplitudes $\gamma_n^{(0)}$:

$$|\psi_\mathrm{Initial}\rangle = \sum_n \gamma_n^{(0)} |n\rangle, \quad (2)$$

perturbation theory can be used to calculate the state of the system at later times, taking into account the action of the interacting part of the Hamiltonian. The probability amplitude of each state at a later time is given by [4]:

$$\gamma_k(t) = \sum_{m=0}^{\infty} \gamma_k^{(m)}(t), \quad (3)$$

where the $\gamma_k^{(m)}$'s are given by:

$$\mathrm{i}\hbar\frac{d}{dt}\gamma_k^{(z+1)}(t) = \sum_n \langle k|\hat{H}_I|n\rangle e^{\mathrm{i}(E_k-E_n)t/\hbar}\gamma_n^{(z)}(t). \quad (4)$$

The zeroth term ($\gamma_k^{(0)}$) is defined by our initial condition, the 1$^\mathrm{st}$ term ($\gamma_k^{(1)}$) accounts for the interaction Hamiltonian to first order (and the $z^\mathrm{th}$ term to $z^\mathrm{th}$ order).

The initial condition ($\gamma_m^{(0)}$s) is unit amplitude in the state described by a single forwards propegating photon and a ground state atom and zero in all others

$$\gamma_m^{(0)} = \delta_m^n, \tag{5}$$

$$|n\rangle = |g1_f\rangle. \tag{6}$$

After an infinite time ($t = \infty$) the MS will have decayed to the ground state resulting in amplitudes in different photonic modes, corresponding to reflection/transmission/loss coefficients. These coefficients are found by summing all of the $\gamma_k^{(z)}(\infty)$ terms, each of which can be found by equ.(4) and is related to the integral over time of the previous term.

The interaction Hamiltonian can only couple two states together if the MS part of the two states is different. Since the MS is both initially and finally in its ground state an even number of $\hat{H}_I$'s must have been applied in total. Thus all of the $\gamma_k^{(n)}(\infty)$ terms with $n$ odd are zero. Each term of equ.(3) is given by the time integral of the previous term. To keep the problem tractable we require that the $+C$ terms introduced by each of these integrals are zero. This is achieved by manipulating the initial time and introducing an infinitesimal.

First, one imposes the initial condition at $t = -\infty$, equivalent to assuming a steady state ("long photon" limit). Second it is required that the interaction Hamiltonian very slowly "turns off" at times approaching $\pm\infty$ as is done in [5, 6]. To achieve this one uses a modified Hamiltonian $\hat{H}_I'(t) = \hat{H}_I e^{-q|t|}$, where $q$ is some positive infinitesimal we will set to zero almost immediately. It is physically intuitive that turning off the interaction infinitely slowly at times approaching $t = \pm\infty$ should have no effect on the result. The purpose of this term and $-\infty$ limit is that:

$$\lim_{q \to 0} \int_{-\infty}^{T} e^{i(a-b)x - q|x|} dx = \frac{e^{i(a-b)T}}{i(a-b)} = \text{anti-derivative}\left[e^{i(a-b)x}\right], \tag{7}$$

i.e. including this infinitesimal $q$ and infinite limit produces the anti-derivative with no $+C$ terms. The ultimate consequence of this will be that each perturbation order ($\gamma_k^{(z)}(\infty)$) can be written as a lower order perturbation multiplied by a constant.

Using this one can integrate equ.(4) to find the amplitudes for being in the ground state with a single photon in mode $p$, $|g1_p\rangle$, given by $\gamma_{g1_p}^2(\infty)$, $\gamma_{g1_p}^4(\infty)$ ... etc. A pattern emerges:

$$\gamma_{g1_p}^{(2)}(\infty) = \frac{2\pi i \delta(E_{g1_p} - E_{g1_f}) \langle g1_p| \hat{H}_I |e0\rangle \langle e0| \hat{H}_I |g1_f\rangle}{(E_e - E_{g1_f})} \tag{8}$$

$$\gamma_k^{(z+2)}(\infty) = \gamma_k^{(z)}(\infty) \sum_j \frac{\langle e0| \hat{H}_I |g1_j\rangle \langle g1_j| \hat{H}_I |e0\rangle}{(E_e - E_{g1_f})(E_g + E_j - E_{g1_f})}. \tag{9}$$

The sum over $j$ includes all photonic modes. Substituting these results into equ.(3) and using the geometric series identity produces:

$$\gamma_k(\infty) = \gamma_k^{(0)} + \frac{2\pi i \delta(E_k - E_{g1_f}) \langle k| \hat{H}_I |e0\rangle \langle e0| \hat{H}_I |g1_f\rangle}{(E_e - E_{g1_f})} \left[1 - \sum_j \frac{\langle e0| \hat{H}_I |g1_j\rangle \langle g1_j| \hat{H}_I |e0\rangle}{(E_e - E_{g1_f})(E_g + E_j - E_{g1_f})}\right]^{-1}. \tag{10}$$

This includes all infinity of the perturbation orders so it is exact to within the approximations in the Hamiltonian. The inclusion of all these increasingly long chains can be thought of in terms of summing over all of the possible paths taking a photon from the initial state to the final state. A cartoon of this is shown in fig.1.

We now evaluate the terms of this equation, beginning with the factor in square brackets.

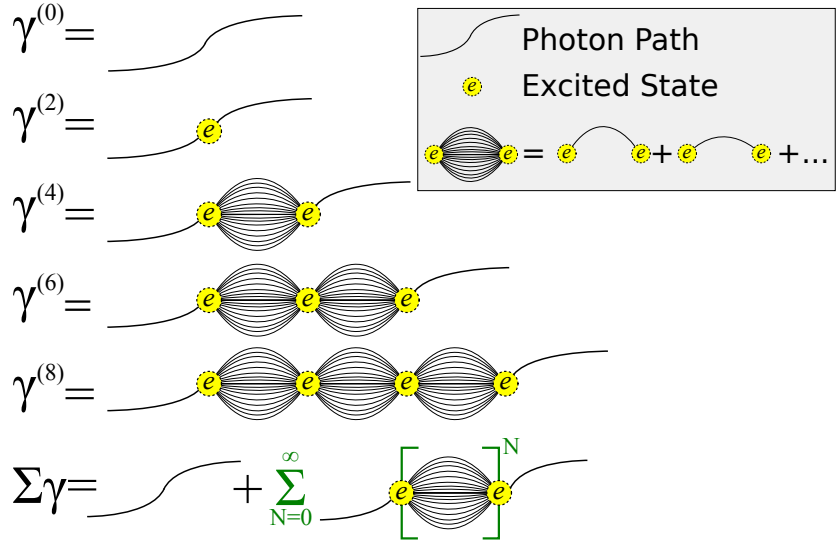

FIG. 1: The amplitude of the photon arriving in state $k$ is given by the sum over every path the photon could take to get to that state. In this case the paths which involve scattering from the dipole zero ($\gamma^{(0)}$), one ($\gamma^{(2)}$), two, three and four times are shown. Each time the single excitation in the system is in a photonic mode it is depicted with a line, when it enters the atom/QD and lifts it to the excited state it is represented by a circle. Multiple lines indicate a sum over the possible paths.

### 1. Evaluation of "[]" factor

Writing this as:

$$[]_{\text{Factor}} = \frac{1}{1-\mathcal{G}} \qquad (11)$$

we can initially concentrate on only $\mathcal{G}$:

$$\mathcal{G} = \sum_j \frac{\langle e0|\hat{H}_I|g1_j\rangle \langle g1_j|\hat{H}_I|e0\rangle}{(E_e - E_{g1_f})(E_g + E_j - E_{g1_f})}. \qquad (12)$$

This term expresses the amplitude with which the excitation can leave the atomic excited state and then be re-absorbed by the atom, including all photon modes. It is represented diagrammatically in fig.2.

The evaluation of this factor proceeds by using a series of Green's function identities to simplify all of the integrals that are implicit in the sum over all photonic modes.

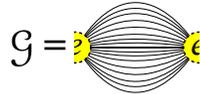

FIG. 2: Cartoon illustrating the role of the term $\mathcal{G}$.

First we substitute in the equation for $\hat{H}_I$ and eliminate the terms that must be zero considering the atomic part of the states and the raising/lowering operators. One can choose to perform the sum over all possible photon states in the position basis, describing each photon state, $|1_{\mathbf{R},\omega,\mathbf{p}}\rangle$, in terms of position, $\mathbf{R}$, frequency, $\omega$ and polarisation, $\mathbf{p}$.

The sum over all photon states ($\sum_j$) is replaced with the integral over position and frequency (as they are continuous) and the sum over polarisations.

The bosonic field operators, $\hat{\mathbf{f}}$, definitionally obey the commutation relation, $[\hat{f}_a(\mathbf{r},\omega), \hat{f}_b^\dagger(\mathbf{r}',\omega')] = \delta(\mathbf{r}-\mathbf{r}')\delta(\omega-\omega')\delta_a^b$, where $a,b$ denote the vector components of the vector-operators [7]. This implies that

$$\sum_{\mathbf{p}} \langle A1_{\mathbf{R},\omega,\mathbf{p}}|\hat{\mathbf{f}}^\dagger(\mathbf{r}'',\omega'')|A0\rangle = \delta(\mathbf{R}-\mathbf{r}'')\delta(\omega-\omega'')\mathbf{I}, \tag{13}$$

where $\mathbf{I}$ is the identity tensor (acting on polarisation) and $A$ is some atomic state. One finds the innermost integrals (those arising from the Hamiltonian) are removed by these $\delta$-functions:

$$\mathcal{G} = -\iint \frac{\mathbf{d}\cdot\mathbf{G}(\mathbf{r},\mathbf{R},\omega)\sqrt{\frac{\hbar\text{Im}[\epsilon(\mathbf{R},\omega)]}{\pi\epsilon_0}}\mathbf{d}^*\cdot\mathbf{G}^*(\mathbf{r},\mathbf{R},\omega)\sqrt{\frac{\hbar\text{Im}[\epsilon(\mathbf{R},\omega)]}{\pi\epsilon_0}}}{(E_e - E_{g1_f})(E_g + \hbar\omega - E_{g1_f})} d^3\mathbf{R} d\omega. \tag{14}$$

Light is reciprocal, which implies $\mathbf{G}(\mathbf{r},\mathbf{R},\omega) = \mathbf{G}^T(\mathbf{R},\mathbf{r},\omega)$. Also note that $\mathbf{d}^*\cdot\mathbf{G}^{*T} = \mathbf{G}^*\cdot\mathbf{d}^*$.

$$\mathcal{G} = -\iint \frac{\hbar\text{Im}[\epsilon(\mathbf{R},\omega)]\mathbf{d}\cdot\mathbf{G}(\mathbf{r},\mathbf{R},\omega)\cdot\mathbf{G}^*(\mathbf{R},\mathbf{r},\omega)\cdot\mathbf{d}^*}{\pi\epsilon_0(E_e-E_{g1_f})(E_g+\hbar\omega-E_{g1_f})} d^3\mathbf{R} d\omega \tag{15}$$

Now use the identity $\int \text{Im}[\epsilon(\mathbf{R},\omega)]\mathbf{G}(\mathbf{r},\mathbf{R},\omega)\cdot\mathbf{G}^*(\mathbf{R},\mathbf{r}',\omega)d^3\mathbf{R} = \text{Im}[\mathbf{G}(\mathbf{r},\mathbf{r}',\omega)]$ [1].

$$\mathcal{G} = -\int \frac{\mathbf{d}\cdot\text{Im}[\mathbf{G}(\mathbf{r},\mathbf{r},\omega)]\cdot\mathbf{d}^*}{\epsilon_0\pi(E_e-E_{g1_f})(E_g/\hbar+\omega-E_{g1_f}/\hbar)} d\omega \tag{16}$$

The energy of the input state ($E_{g1_f}$) is the ground state energy of the atom plus the energy of the input photon ($E_g + \omega_f\hbar$), we substitute this in and use the relation $1/(a-b) = -i\pi\delta(a-b) + \mathcal{P}/(a-b)$ with $\mathcal{P}$ the principal value [8].

$$\mathcal{G} = -\frac{-i\mathbf{d}\cdot\text{Im}[\mathbf{G}(\mathbf{r},\mathbf{r},\omega_f)]\cdot\mathbf{d}^*}{\epsilon_0(E_e-E_{g1_f})} - \int \frac{\mathcal{P}\mathbf{d}\cdot\text{Im}[\mathbf{G}(\mathbf{r},\mathbf{r},\omega)]\cdot\mathbf{d}^*}{\epsilon_0\pi(E_e-E_{g1_f})(\omega-\omega_f)} d\omega \tag{17}$$

The result of the integral in the second term is real and contributes a shift in the resonant energy of the optical transition (Lamb shift). This shift is not the main concern of this work and will not be studied in detail. Setting the limits of this integral to $\pm\infty$ (integrating over all frequencies) it can be simplified using the Krammers Kronig relation [6] giving:

$$\mathcal{G} = -\frac{-i\mathbf{d}\cdot\text{Im}[\mathbf{G}(\mathbf{r},\mathbf{r},\omega_f)]\cdot\mathbf{d}^* + \mathbf{d}\cdot\text{Re}[\mathbf{G}(\mathbf{r},\mathbf{r},\omega_f))]\cdot\mathbf{d}^*}{\epsilon_0(E_e-E_{g1_f})}. \tag{18}$$

The entire "square bracket factor" was given by $1/(1-\mathcal{G})$. It can now be expressed:

$$[]_{\text{Factor}} = \frac{E_e - E_{g1_f}}{E_e - E_{g1_f} + \mathbf{d}\cdot\text{Re}[\mathbf{G}(\mathbf{r},\mathbf{r},\omega_f))]\cdot\mathbf{d}^*/\epsilon_0 - i\mathbf{d}\cdot\text{Im}[\mathbf{G}(\mathbf{r},\mathbf{r},\omega_f)]\cdot\mathbf{d}^*/\epsilon_0}. \tag{19}$$

With this result equ.(10) now reads:

$$\gamma_p(\infty) = \gamma_p^{(0)}(0) + \frac{2\pi i\delta(E_{g1_p}-E_{g1_f})\langle g1_p|\hat{H}_I|e0\rangle\langle e0|\hat{H}_I|g1_f\rangle}{E_e - E_{g1_f} + \mathbf{d}\cdot\text{Re}[\mathbf{G}(\mathbf{r},\mathbf{r},\omega_f))]\cdot\mathbf{d}^*/\epsilon_0 - i\mathbf{d}\cdot\text{Im}[\mathbf{G}(\mathbf{r},\mathbf{r},\omega_f)]\cdot\mathbf{d}^*/\epsilon_0}. \tag{20}$$

The $\langle e0| \hat{H}_I |g1_f\rangle$ terms are $\sqrt{\text{coupling}}$ rates between the excited state and photon $f$. They are given by [9, 10]:

$$\langle e0| \hat{H}_I |g1_f\rangle \propto \sqrt{\frac{\omega_{g1_f}\hbar}{2\epsilon_0}}\mathbf{d}^* \cdot \mathbf{E}_f(\mathbf{r}, E_{g1f}). \tag{21}$$

We assume a single mode waveguide, where the propagation direction, together with the energy $E_{g1f}$, uniquely selects for a particular waveguide mode. The constant of proportionality in equ.(21) depends on how the modes are indexed [10]. In the specific case where the modes are from a discrete set indexed by $k$-vector the constant of proportionality can be set to one.

Finally we assume that the scattered photon is detected by some device that has a finite energy resolution, and integrate equ.(20) across some small interval of energies that are unresolvable from one another for the detector. The purpose of this is to remove the $\delta$-function. One might worry at this point that choosing to integrate over an energy window, instead of an angular frequency window or $k_x$ interval appears to affect the outcome of the integral. For example $\hbar\delta(E_{g1_p} - E_{g1_f}) = \delta(\omega_{g1_p} - \omega_{g1_f})$.

However, there is also the choice of how one indexes the modes which affects how they must be normalised and consequently determines the proportionality factor in equ.(21). We find that any consistent choice (i.e. the modes are indexed by the same parameter that is integrated over) produces the same result as any other consistent choice:

$$\int_{\text{Modes}\approx p} 2\pi\mathrm{i}\delta(E_{g1_p} - E_{g1_f})\langle g1_p| \hat{H}_I |e0\rangle \langle e0| \hat{H}_I |g1_f\rangle = \frac{\mathrm{i}a\omega_{g1_f}}{2|v_g|\epsilon_0}\mathbf{d}\cdot\mathbf{E}_p^*(\mathbf{r})\mathbf{d}^*\cdot\mathbf{E}_f(\mathbf{r}), \tag{22}$$

where the $\mathbf{E}$ fields are given for the Bloch parts of modes indexed discretely by their $k$-vector, and $v_g$ is the group velocity, $v_g = d\omega/dk$.

With the numerator part of equ.(20) known we can move onto the denominator. To solve this part we substitute the Greens function as $\mathbf{G}_{\text{Tot}} = \mathbf{G}_{\text{WG}} + \mathbf{G}_{\text{Loss}}$. The Green's function for the PhCWg ($\mathbf{G}_{\text{WG}}$), is given in [9]. We also set the detuning to $\Delta$ (including the Lamb-shift term).

This produces our final expression, where $\gamma_{\sim p}$ represents the probability amplitude integrated over some tiny window surrounding the state $\gamma_p$ (including only modes in the same direction):

$$\gamma_{\sim p}(\infty) = \gamma_p^{(0)}(0) - \frac{\mathbf{d}\cdot\mathbf{E}_p^*(\mathbf{r})\mathbf{d}^*\cdot\mathbf{E}_f(\mathbf{r})}{0.5\left[|\mathbf{d}\cdot\mathbf{E}_f(\mathbf{r})|^2 + |\mathbf{d}\cdot\mathbf{E}_b(\mathbf{r})|^2\right] + \frac{2v_g}{a\omega}\left[\mathbf{d}\cdot\mathbf{G}_{\text{Loss}}\cdot\mathbf{d}^* + \mathrm{i}\hbar\epsilon_0\Delta\right]}. \tag{23}$$

Equ.(23) can be cast into a reflection coefficient by setting $p = b$ (i.e. backwards mode) and $\gamma_b^{(0)}(0) = 0$ (initially no backwards propagating amplitude). For a transmission coefficient one instead sets $p = f$ and $\gamma_p^{(0)}(0) = 1$ (for the photon initially in the forwards mode).

### B. Emission

From Fermi's golden rule the decay rate of the excited system in the forwards direction is given by [4]:

$$\Gamma_f = \frac{2\pi}{\hbar}\rho|\langle g1_f| \hat{H}_I |e\rangle|^2, \tag{24}$$

with $\rho$ the density of states.

Substituting the results from the previous section, beginning with equ.(21), and setting the density of states with respect to energy to $\rho = a/(2\pi\hbar|v_g|)$ one finds:

$$\begin{aligned}\Gamma_f &= \frac{a\omega}{2|v_g|\hbar\epsilon_0}|\mathbf{d}^* \cdot \mathbf{E}_f(\mathbf{r})|^2, \\ \Gamma_b &= \frac{a\omega}{2|v_g|\hbar\epsilon_0}|\mathbf{d}^* \cdot \mathbf{E}_b(\mathbf{r})|^2.\end{aligned} \quad (25)$$

## II. CHANGING THE TARGET DIRECTIONALITY

As discussed in the main text the area of the waveguide suitable for unidirectional coupling increases when we consider dipoles other than circular, with the ideal case being dipoles that are optimised to the polarisation of the waveguide mode at their location.

The usable area always increases with this new freedom, but the size of the increase depends on the exact assumptions made. For example if we are given complete freedom to modify the dipole then a directional contrast of unity ($|D| = 1$) is possible everywhere in the waveguide except on lines of zero area. At the opposite extreme a circular dipole only has $|D| = 1$ at point-like locations in the waveguide with zero area.

In the figure we consider some intermediate assumptions. For each given directionality, for the waveguide mode described in the letter, we calculate the area where a circular dipole achieves the given modulus value of directionality (or higher). We similarly calculate the area for a dipole that can be tuned within limits. The ratio of these areas (tunable / circular) is plotted in fig. 3. We consider two tunable dipoles, one that (like in the main text) is constrained to have $S_3 \geq 1/\sqrt{2}$ and another where we assume a higher degree of tuneability so that $S_3 \geq 1/2$.

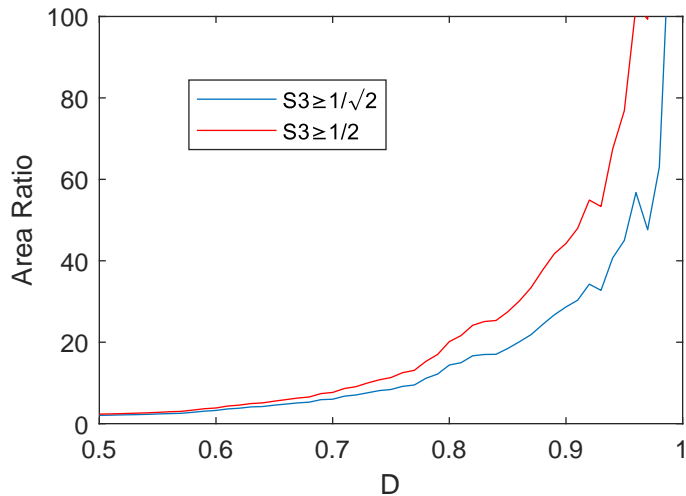

FIG. 3: The area suitable for directionality greater than or equal to $|D|$ with a tunable dipole, divided by the area with a circular dipole. The noise is a product of the simulation's finite resolution (pixels of length $a/48$).

One can see that the increase in the usable area is higher at higher directionalities. Only pointlike locations in the waveguide support a directionality of $|D| = 1$ for circular dipoles, while $|D| = 1$ is achievable in a finite area when the dipoles is somewhat tunable. Consequently, the ratio diverges at $|D| = 1$. At the other end of the plot range about 1/3rd of the waveguide area supports $|D| > 0.5$ for circular dipoles, which increases to about 2/3rds when the dipoles are tunable, meaning that the ratio here is 2.

As discussed in the main text the usable area increases by about 30 times for a target directionality of $D = 0.9$. We highlight this value as this level of directionality is the regime of current QD experiments in photonic crystal waveguides [11].

## III. A NOTE ON THE ORTHOGONALITY OF DIPOLES

In this section we will discuss a misconception that held back the work described in the main text. By explicitly stating this misconception and the errors it results in we hope that we may help other researchers avoid a similar error.

*The Mistaken view*: The quantum system starts in one of two excited states that are orthogonal. However, during the emission process the decays from each of the two excited states are described by dipoles that are not orthogonal (fig. 4(b) of the main text). Surely the dipoles must also have to be orthogonal, as unitary evolution cannot evolve orthogonal quantum states to non-orthogonal ones.

The key logical error is equating the orthogonality of quantum states in Hilbert space with complex vectors in real space. On the emitter side there is no reason to assume that $\langle e_1|e_2\rangle = 0$ must imply that $\mathbf{d}_1^* \cdot \mathbf{d}_2 = 0$, where $\mathbf{d}_{1,2}$ are the dipoles related to the decay processes $|e_{1,2}\rangle \to |g_{1,2}\rangle$.

A related misconception equates superpositions of states in Hilbert space with linear combinations (classical superpositions) of dipole vectors. Once again relating the two is not justified by any theoretical considerations, and can lead to provably incorrect results.

Given the right waveguide location two non-orthogonal dipoles can be distinguished perfectly, as one only radiates forwards, the other only backwards. fig.4(b). Once again there is no issue, $\langle f|b\rangle = 0$ does not require $\mathbf{d}_1^* \cdot \mathbf{d}_2 = 0$. However one might still be unsettled, given that two planewave photons with non-orthogonal polarisations cannot be distinguished perfectly why can two dipoles described by the same complex vectors?

The difference is that, unlike the photon polarisation the dipole does *not* describe a quantum state. It is associated with the *operator* that changes the state of the MS and related light field [equ.(2) in the letter]. Being described by an operator means that a dipole has much more in common with a *polariser* than it does with a *polarisation*. A linear polariser that is somehow prepared in a quantum superposition of two orientations is not a circular polariser, instead it becomes entangled with passing photons. This is analogous the the quantum protocols described in the text where the QD state (determining the dipole) becomes entangled with passing photons.

---



\end{document}